\documentclass[twocolumn, showpacs]{revtex4} 

\usepackage{amsmath}
\usepackage{graphicx}

\begin{document}
\title{ A Universal Origin of Information Accumulation in Nature}  
\author{			Keiichi Akama}
\affiliation{	Department of Physics, Saitama Medical University,
 			 Saitama, 350-0495, Japan}
\date{\today}

\pacs{01.70.+w, 05.70.-a, 89.70.Cf , 89.70.-a }



\begin{abstract}	
To account for the origin of information accumulation in nature 
	despite the entropy-increase law,  
	we advocate a universal mechanism 
	due to competition/selection of general composite entities, from simple to complex. 
To confirm its universality, we show that even simplest composites such as an atom and a molecule  
	are subject to this mechanism and accumulate information.
\end{abstract}

\maketitle

Thermodynamics or statistical physics indicates that entropy increases in closed systems
	and nature should tend to disorder as a whole \cite{Ref1}.
On the other hand, we observe intensive accumulation of information in nature. 	
It is the tremendous amount of information accumulated there 
	that enables the marvels of the cosmos, the miracles of life, 	
	the profundity of human nature, etc. 	
They do not contradict each other, since entropy increases globally in large closed systems,   
	while information is accumulated locally in specific open systems. 
Nevertheless, the physical laws cannot provide any fundamental explanation 
	for the latter feature, which is so prominent in nature.
We want to know the universal origin of information accumulation 
	to complement the entropy-increase law. 
For this purpose, we advocate an accumulation mechanism, 
	which is based on competition for existence among general composite entities 
	and the resultant selection \cite{Akama}.

All the entities in nature
	are fundamental fields (photons, electrons, quarks, etc.\ \cite{elementary}) 
	or their simple or multiple composites \cite{composite}
	(atoms, molecules, stones, mountains, animals, 
	societies, stars, galaxies, etc.\ \cite{braneworld}). 
The composites are formed via selection out of competing possibilities	
	according to natural laws and given conditions. 	
We call the function {\it the physical selector}. 
The laws and the conditions are the contents of the information 
	which enables existence of the composites.
According to Bateson, the information is ``a difference which makes a difference" \cite{Bateson}.
We call the made difference its {\it effect} or {\it goal}. 
Selection makes a difference. 
Hence, {\it selection for an effect makes information for the effect}.
Here, the effect is the compositeness. 
Furthermore, 
	thus formed composites 
	interact with each other, compete for their existence,  
	and are selected via real processes, 
	so that the information is accumulated. 
We call this function {\it the real selector}. 	
If the composites are living beings, 
	it is nothing but Darwin's natural selection \cite{Darwin} 
	in its literal meaning \cite{literal}. 
Aliveness is a form of compositeness.
The information is so organized to form algorithms for compositeness.
Larger multiple composites have capacity for more refined information. 
Superior information contributes more to the composite's duration, and is selected  	
	so as to endure longer together with the composite, 
	whereas inferior information disappears together with the lost composite. 
Thus, {information accumulates information by virtue of itself}.

In the long course of successive self-accumulation of information,
	various strategies and functions for duration are selected and accumulated as information
	(e.g.\ co-operation, proliferation, growth) \cite{examples}. 
Information accumulated for intermediate goals (e.g.\ instincts, emotions)  
	would develop a profusion of activities as by-products (e.g.\ arts, sciences).
In particular, information for the functions to select something 
	accelerates information accumulation. 
We call the functions {\it the inner selectors}. 	
For example, real selection of individuals is inner selection for species,	
	animals select responses to create information in information processing in their brain, and so on. 
Thus, various strategies are selected and accumulated as information, 
	including self-organization, information coding, repairing, 
	self-replication, heredity, cognition, cultures, etc. 
We are astonished by ingenious algorithms for duration of some super-multiple composites 
	such as living organisms. 	
Who wrote them, and how?
They are, we suppose, the results of successive information accumulation 
	driven by competition/selection among general composites, from simplest to complex, 
	though details are yet to be investigated.

In short, {\it composites are formed with information, compete for existence, 
	are sifted by various selectors, and accumulate information successively,
	until they are lost in competition}. 
Information in nature is made by some selection 
	for duration of some composites or for some by-products. 	
As for complex composites like organisms, the mechanism is working rather obviously \cite{Darwin}.     	
Here, we claim that it is {\it universal for all the composites} from hadrons to celestial structures, including life.
Otherwise, it cannot complete itself, since the interactions are borderless.
Then, what is urgent for us is to confirm it for simplest composites as well. 	
In the following, we see that even an atom, a molecule, etc.\ compete for existence, undergo selection,
	and accumulate information, 
	while an extended entropy-increase law holds on average.

A hydrogen atom is composed of a proton and an electron with electric force.
Nature selects, according to quantum laws and conditions, 
	a finite number (say $n^-$) of negative-energy states 
	(with zero energy at infinity) to form a composite, the atom. 
Whereas, positive-energy states fail to form composites, 
	and the other continuous-energy states are deselected. 
The laws and the conditions are the contents of the information for existence of the atom.
The existence depends also on interactions with other composites.
The information resides over the environment. 
Here we assume that the system is in a cubic container with edge length $L$
	in a heat reservoir of temperature $T$. 
It specifies the distribution of the environmental disturbances, 
	and hence specifies the probability distribution $p $ of the state $i$ of the system:
\begin{eqnarray}&&
	p_i =e^{- E_i/ k{}T} /Z \ \ {\rm with\ \ }Z\equiv {\sum}_i e^{- E_i/ k{}T},
   \label{p_i=} 
\end{eqnarray}
where $E_i$ is the energy of the state $i$, and
	$ k{}$ is the Boltzmann constant. 
Though the atom has neither temperature nor thermodynamic entropy,
we can define the Shannon entropy $ S\equiv -{\sum}_i p_i \ln p_i $ \cite{Shannon}.
Let us consider the change of the distribution $p$ from $p ^{\rm in}$ to $p ^{\rm fi}$,
	and denote $ X^{\rm in(fi)}\equiv X|_{p =p ^{\rm in(fi)}} $ for $X$ concerned with $p$. 
Then, the amount of information accumulated in the change 
	is given by the decrease $-\Delta S\equiv S^{\rm in }-S^{\rm fi }$ of $S$.

Let us examine the entropy-increase law. 
We assume that only $T^{\rm in}$ and $T^{\rm fi}$ are visible in the heat reservoir, 
	and the invisible processes finally yield irreversibility of the final state.
Then, the change of the environmental entropy associated with the system is given 
	by $\Delta \tilde S\equiv {\sum}_i (p_i^{\rm fi}- p_i^{\rm in}) \ln p_i ^{\rm fi}$. 	
In fact, in terms of (\ref{p_i=}), it becomes
\begin{eqnarray}&&
	\Delta\tilde S= Q/kT^{\rm fi}\ \ {\rm with\ \ } 
	Q\equiv \langle E \rangle ^{\rm in } -\langle E \rangle ^{\rm fi },
   \label{DStilde}
\end{eqnarray}
which reproduces the thermodynamic relation (with averaged energy $\langle E \rangle $).
Then, the change $\Delta S_{\rm tot} \equiv \Delta S+\Delta \tilde S $ 
	of the total entropy is given by 
\begin{eqnarray}&&
	\Delta S_{\rm tot}= {\sum}_i p_i ^{\rm in} \ln (p_i  ^{\rm in}/p_i ^{\rm fi})
	=D(p ^{\rm in}||p ^{\rm fi})\ge0,
   \label{DStot}
\end{eqnarray}
where $ D(q ||q') \equiv {\sum}_i q_i \ln (q_i / q'_i)$ (for the probability distributions $q $ and $q' $)
	is the Kullback-Leibler divergence, which is known to be non-negative \cite{KL}. 
Eq.\ (\ref{DStot}) is an extended entropy-increase law in terms of the Shannon entropy.
It is, however, broken in some specific processes, 
	since, unlike in the statistical limit, $\Delta S_{\rm tot}$ fluctuates by 
	$\sigma_E^{\rm in}|1/ k T^{\rm in}-1/ k T^{\rm fi}| $,
	where $\sigma_E^{\rm in}$ is the fluctuation of $\langle E \rangle^{\rm in}$.
The fluctuation would trigger information accumulation due to competition/selection in some specific composites
	despite global increase of total entropy.

Then, we consider the amount of information stored by the atom for its existence in the environment. 
Let $p^+$ ($p^-$) be the conditional probabilities for unbound (bound) states,
	$Z^\pm$ be the partition functions for $ p^{\pm}$ ($Z=Z^-+Z^+$), 
	and $S^\pm$ be the Shannon entropies with $ p^{\pm}$.
Then, $I\equiv S^+-S^-$ indicates the information storage of the atom. 
The higher states are approximated by the ideal gas of an electron and a proton, so that
\begin{eqnarray}
	Z^+ = (L^2m_{\rm e} k{} T/2\pi \hbar^2)^{3/2}(L^2m_{\rm p} k{} T/2\pi \hbar^2)^{3/2},    
\label{Z_+=}
\end{eqnarray}
where $m_{\rm e}$ is the mass of an electron, $ m_{\rm p} $ is the mass of a proton, 
	and $\hbar=h/2\pi$ with the Planck constant $h$.
The lower states are well approximated by those with infinite volume, 
	and $ Z^-$ is approximately given by 
\begin{eqnarray}
	Z^-={\sum}_{n=1}^{n^{\rm -}}n^2 e^{ E^*/n^2kT}
	(L^2m_{\rm p} k{} T/2\pi \hbar^2)^{3/2},
\label{Z_-=}
\end{eqnarray}
	where $E^*$ is the ionization energy.
Because $Z^+$ ($Z^-$) dominates $Z$ at high (low) temperatures, 
	the Shannon entropy $S$ rapidly increases from $S^-$ to $S^+$ with increasing $T$
	around the cross-over temperature $ T^{\rm c}=T|_{Z^-=Z^+}$ \cite{Akama}.
Then, it is appropriate to estimate the information storage $ I$ at $ T=T^{\rm c}$.
Below $T^{\rm c}$, the ground state dominates $Z$. 
In terms of (\ref{Z_+=}) and (\ref{Z_-=}), we obtain  
\begin{eqnarray}&&
	T^{\rm c} 
	= 2E^*/3k{} \eta^{\rm inv} (L^2 m_{\rm e} E^* / 3\pi\hbar^2),    
\label{T_c=}
\\&&
	I^{\rm c}\equiv I|_{T=T^{\rm c}} =3/2+ E^*/k{} T^{\rm c},
\label{DeltaI_c=}
\end{eqnarray}
where $x=\eta^{\rm inv}(y)$ is the inverse of $y=\eta(x)\equiv xe^x$.
The fluctuation $\sigma$ of $ I^{\rm c}$ is given 
	by $\sigma^2=\sigma_{ +}^2+\sigma_{ -}^2$,
where $\sigma_{ \pm}$ are the fluctuations of $S^\pm$. 
With (\ref{Z_+=}) and (\ref{Z_-=}), we have
	$\sigma_{ +}^2=3$ and $\sigma_{ -}^2=3/2$.
For example, for $L=10^{-7} $m, we have $ T^{\rm c}= 1.07\times10^4$K 
	and $ I^{\rm c}\pm\sigma=(23.5\pm3.1) $bit.
This is taken as the information amount of the algorithm 
	by which the atom composes itself and endures the disturbances from the heat reservoir \cite{intention}.

The atoms have a repair function as their intrinsic information. 
Suppose that the environment has extra sources of disturbances and they excite the atom. 
The excited atom would easier be decomposed by subsequent disturbances with less energy,
	and, hence, its information is disadvantageous for duration.
The atom, however, could spontaneously emit photons to recover its original securer state. 
This is taken as an inner selection by the atom \cite{intention}.  
Now we inquire the amounts of information in repairing.
Let $p_{ji}$ ($i\not=j$) be the transition probability 
	from the state $i$ to the state $j$ due to the extra disturbances.
Then, the probability $p' $ of the excited state $i$ is given by 
\begin{eqnarray}&&
	p'_i=p_i-Z^{-1}{\sum}_{j=1}^{\infty} (p_{ji}e^{-E_i/kT}- p_{ij}e^{-E_j/kT}),\ \ \ \ 
\label{p'=}
\end{eqnarray}
which is non-canonical, and (\ref{DStilde}) does not hold for $p'$. 	
In the excitation, the increase of Shannon entropy, i.e.\ the loss of information becomes
\begin{eqnarray}&&
	\Delta S^{\rm ex}			
	=\Delta \langle E \rangle/kT    
	-Z^{-1} {\sum}_{i=1}^{\infty}e^{-E_i/kT}r_i \ln r_i \ \ \ \ \ \ \
\label{DHex}
\end{eqnarray}
where 	$ r_i \equiv 1-{\Sigma}_{j=1}^\infty( p_{ji}- p_{ij}e^{(E_i-E_j)/kT})$,
	and $\Delta \langle E \rangle\equiv \langle E \rangle'- \langle E \rangle $ 
	($\langle E \rangle' $ is the energy averaged with $p' $),
while in the recovery, it changes just by $-\Delta S^{\rm ex}$.
The definition of $ \Delta \tilde S $ is still relevant for the non-canonical distribution $p'$ in (\ref{p'=}). 
Therefore, the entropy-increase law (\ref{DStot}) still holds (on average).
In fact, the total entropy $ S_{\rm tot}$ increases both in the excitation ($p ^{\rm in}=p $, $p ^{\rm fi}=p' $) 
	and in the recovery ($p ^{\rm in}=p' $, $p ^{\rm fi}=p $), respectively, by 
\begin{eqnarray}&&\hskip-5pt
	\Delta S_{\rm tot}^{\rm ex}
	=-Z^{-1}{\sum}_{i=1}^{\infty} e^{-E_i/kT} \ln r_i = D(p ||p')\ge0, \ \ \ \ \label{Dex}\ \ 
\\&&\hskip-5pt 
	\Delta S_{\rm tot}^{\rm rec}
	=Z^{-1}{\sum}_{i=1}^{\infty} e^{-E_i/kT}r_i \ln r_i= D(p' ||p) \ge0.\ \ \ \ \label{Drc}
\end{eqnarray}
In particular in the recovery, eq.\ (\ref{DStilde}) also holds since the $ p ^{\rm fi}$ is canonical.
The disturbances may come from formations or recoveries of other composites, 
	and the emitted photons may affect other composites, or they may directly collide.	
They are in struggle for existence, 
	and, as a result of the selection, the information is accumulated with the survivors.

The atoms further accumulate information by forming molecules. 
The atoms adapt their states to form the structures, and get advantages for their own duration~\cite{intention}. 
It is an inner selector of the atom.
At the same time, it is the composing information of the molecule.
It is a general characteristic of information to have different meanings according to users.
Let us consider two hydrogen atoms in the container specified above. 
The partition function for unbound states is approximated by 
\begin{eqnarray}&&
	Z_{\rm 2H}= Z_{\rm H}^2/2 \ \ {\rm with\ \ }	
	Z_{\rm H}\equiv (L^2m_{\rm p} k{}T/2\pi\hbar^2)^{3/2}, 
	\ \ \ \ \ 
   \label{Zmol+}
\end{eqnarray}
while that for bound states, i.\ e.\ for a molecule, is 	
\begin{eqnarray}&&
	Z_{\rm H_2}=e^{E^*_{\rm m}/k{}T} 	
	I_{\rm_H} (L^2m_{\rm p}/\pi) ^{3/2} (k{}T/ \hbar^2)^{5/2}q_{\rm m}(T),
	\ \ \ \ \ 
   \label{Zmol-}
\end{eqnarray}
where $ E^*_{\rm m}$ is the binding energy of the molecule, $I_{\rm_H}$ is its moment of inertia,
	$ q_{\rm m}(T) \equiv 1+\Sigma_i e^{-\epsilon_i /k{}T}$, 
	and $\epsilon_i $ ($i=1,2,\cdots$) is the vibrational excitation energies.
We estimate their cross-over temperature with $T^{\rm c}_{\rm m}\equiv T|_{ Z_{\rm 2H}=Z_{\rm H_2}}$.
We use $ I^{\rm c}_{\rm m}\equiv ( S_{\rm 2H} - S_{\rm H_2})|_{T=T^{\rm c}_{\rm m}}$
	to estimate the information storage by the atom,
	where $ S_{\rm 2H}$ ($S_{\rm H_2}$) is 
	the Shannon entropy based on $ Z_{\rm 2H}$ ($Z_{\rm H_2}$).
With (\ref{Zmol+})--(\ref{Zmol-}), we have
\begin{eqnarray}&&
	T^{\rm c}_{\rm m} 
	= 2E^*_{\rm m}/k{} \eta^{\rm inv}_{\rm m} 
	(L^6 m_{\rm p}^3 E^*_{\rm m} / 128\pi^3\hbar^2I_{\rm_H}^2), \ \ \ \      
\label{T_cmol=}
\\&&
	I^{\rm c}_{\rm m} 
	=1/2
	+(E^*_{\rm m}-\langle \epsilon \rangle|_{T= T^{\rm c}_{\rm m}})/k{} T^{\rm c}_{\rm m},
\label{I_cmol=}
\end{eqnarray}
where $x=\eta^{\rm inv}_{\rm m}(y)$ is the inverse function of 
	$ y=\eta_{\rm m}(x)\equiv xe^x [q_{\rm m} (2 E^*_{\rm m}/ k{}x)]^2$,
	and $\langle \epsilon \rangle $ is the average of $\epsilon_i$.
The fluctuation $\sigma _{\rm m}$ of $ I^{\rm c}_{\rm m}$ is given by 
	$\sigma _{\rm m}^2=11/2+(\sigma _{\epsilon}/kT)^2$, 
	where $\sigma _{\epsilon}$ is the fluctuation of $\langle\epsilon \rangle $. 
For example, for $L=10^{-7} $m, we have $ T^{\rm c}= 2.67\times10^3$K 
	and $ I^{\rm c}_{\rm m}\pm\sigma _{\rm m}=(28.3\pm3.6) $bit, 
	where we used phenomenological values for $ E^*_{\rm m}$, $\epsilon_i $ \cite{Moore}, 
	and $I_{\rm_H}$ \cite{Hori}.
This is taken as the information amount of the algorithm
	with which the hydrogen molecule composes itself and endures the thermal disturbances,
	as well as that for which the atoms adapt their states to form the molecule for their own duration
	\cite{intention}.

If a composite is formed in some circumstances, 
	it is plausible that many of them are formed, 
	since the conditions for the formation are similarly fulfilled.  
Thus, the composites proliferate in plenty. 
In particular, if existing composites contribute to new formations,
	they would be formed efficiently, and we call it self-proliferation.
The proliferations enhance chances of competition, and hence, of information accumulation.
The group of proliferated composites would form a composite due to some informational connections, if any
	(e.g.\ gases, species).
Let us investigate the information amount of the group composites.
We denote the variables for the group composite with $N$ pieces (atoms or molecules) by those with the suffix $N$.
If $N\ll N_{\rm s}\equiv(L\sqrt{mkT}/h)^3$ ($m$ is the mass of the piece), 
	and if we neglect the small interactions among the pieces,
	we have the partition function $Z_N=Z_1^N/N!$, 
	and hence the Shannon entropy $S_N=NS_1-\ln N!$,
	and the Kullback-Leibler divergence $D_N=ND_1 $.
On the other hand, $S_N$ fluctuates by $\sigma_{S_N} = \sqrt{N}\sigma_{S_1}$,  
	and $D_N$, by $\sigma_{D_N} = \sqrt{N}\sigma_{D_1}$.
The relative importance of fluctuations is suppressed by the factor $\sqrt{N}$, 
	and the statistical physics comes in power.
The entropy-increase law holds more accurately. 
The partition function for $2N$ atoms ($N$ molecules) 
	is given by $ Z_{2N\rm H}=Z_{\rm H}^{2N}/(2N)!$ ($Z_{N\rm H_2}=Z_{\rm H_2}^{N}/N!$). 
We approximate the cross-over temperature $ T^{\rm c}_N$ by $T|_{Z_{2N \rm H}= Z_{N\rm H_2}}$, 
	and the information storage $I^{\rm c}_N$ by $(S_{2N \rm H}- S_{N\rm H_2})|_{T= T^{\rm c}_N }$ 
	with the Shannon entropy $ S_{2N \rm H}$ $ (S_{N\rm H_2})$ 
	based on $ Z_{2N \rm H}$ $ (Z_{N\rm H_2})$.
With (\ref{Zmol+})--(\ref{Zmol-}), 
\begin{eqnarray}&&
	T^{\rm c}_{N} 
	= 2E^*_{\rm m}/k{} \eta^{\rm inv}_{\rm m} 
	(L^6 m_{\rm_H}^3 E^*_{\rm m} / 32\kappa^2\pi^3\hbar^2I_{\rm_H}^2), \ \ \ \      
\label{T_cNmol=}
\\&&
	I^{\rm c}_{N } 
	=N/2+N(E^*_{\rm m}-\langle \epsilon \rangle|_{T= T^{\rm c}_{N }})
	/k{} T^{\rm c}_{N },
\label{I_cNmol=}
\end{eqnarray}
where $\kappa\equiv [(2N)!/N!]^{1/N}$.
Apart from the information due to the neglected connections of the pieces, 
	the information storage is roughly proportional to $N$.
Though the treatments above are semi-classical, 
	quantum properties such as discrete spectra, quantum uncertainty, and 
	exclusive occupations by fermions play essential roles in the competition/selection.
On the other hand, if $N\gtrsim N_s$, 
	it requires fully quantum theoretical treatments with von Neumann entropy \cite{Neumann}.

The group composites further accumulate information 
	by forming liquids, crystals, and other structures 
	with stronger connections among its pieces. 
It depends on how natural laws and conditions indicate. 	
In fact, nature provides a profusion of possibilities.
Let us consider a model of piece trapping in large molecule formation, crystal growth etc. 
Suppose that the trapping potential is given by independent oscillators 
	along the $j$-th spatial axis ($j=1,2,3$)
	with excitation energies $\epsilon_i^{(j)}$ ($i=1,2,\cdots$).
The partition functions for the unbound and trapped states are approximated, respectively, by 
\begin{eqnarray}&&
	Z^+_{\rm t}= L^3 ( m_{\rm t} k{}T/2\pi\hbar^2)^{f/2}, 
   \label{Ztrap+}
\\&&
	Z^-_{\rm t}=e^{E^*_{\rm t}/k{}T} {\Pi}_{j=1}^3 q_{\rm t}^{(j)}(T),
   \label{Ztrap-}
\end{eqnarray}
where $ m_{\rm t}$ is the mass of the piece, 
	$f$ is its degree of freedom in the ideal gas,
	$ E^*_{\rm t}$ is the binding energy,
	and $ q_{\rm t}^{(j)} (T) \equiv 1+\Sigma_i e^{-\epsilon_i^{(j)} /k{}T}$ ($j$=1,2,3).
We estimate the cross-over temperature with $T^{\rm c}_{\rm t}\equiv T_{ Z^+_{\rm t}=Z^-_{\rm t}}$,
	and the information storage 
	with $ I^{\rm c}_{\rm t}\equiv (S^+_{\rm t}-S^-_{\rm t})|_{ T=T^{\rm c}_{\rm t}}$,
	where $ S^\pm_{\rm t}$ is the Shannon entropy based on $ Z^\pm_{\rm t}$.
Then, we have
\begin{eqnarray}&&
	T^{\rm c}_{\rm t} = 2E^*_{\rm t}/f k{} \eta^{\rm inv}_{\rm t} 
	(L^{6/f} m_{\rm t}^3 E^*_{\rm t} / f\pi^3\hbar^2), \ \ \ \      
\label{T_ctrap=}
\\&&
	I^{\rm c}_{\rm t} 
	=f/2+(E^*_{\rm t}-{\sum}_{j=1}^3\langle\epsilon ^{(j)}\rangle|_{T= T^{\rm c}_{\rm t}})
	/k{} T^{\rm c}_{\rm t},
\label{I_ctrap=}
\end{eqnarray}
	where $x=\eta^{\rm inv}_{\rm t}(y)$ is the inverse function of 
	$ y=\eta_{\rm t} (x)\equiv xe^x [\Pi_j q_{\rm t}^{(j)} (2 E_{\rm t} ^*/f k{}x)]^{2/f} $.
In general, larger structures have larger capacity for information and 
	provide more stable environment for the pieces.
The pieces select their states to form the structures and get advantages for their duration \cite{intention}.  
For example, their partners serve as protection barriers against disturbances.
The pieces cooperate for the benefits at the cost of possible self-sacrifice.
It is interesting that living beings often use the survival strategies as atoms and molecules do. 
Thus, we have seen that even the simple composites compete for existence, undergo selection, 
	and accumulate information. 	
The amount is small but sufficient for their duration.

In summary, we advocated the universal information-accumulation mechanism
	due to competition/selection for compositeness. 
It is rather obvious in complex composites like organisms \cite{Darwin}. 
We saw that it works well even in simplest composites. 	
This supports its universality in nature.
If it is universal, natural selection \cite{literal} would be raised from a mechanism in biology 
	to a fundamental law of physics. 
The object of selection is extended from living beings to general composites,
	and its theoretical aim, from the origin of species to that of information accumulation in general. 
It would complement the entropy-increase law in nature,
	though it still requires confirmation in wide classes of composite entities.

The author would like to thank Dr.~T.~Hattori, Dr.~Y.~Kawamura, Dr.~H.~Mukaida, Dr.~S.~Suzuki, Dr.~S.~Wada,
	and Dr.~S.~Yazaki for discussions.


\small
\noindent
{\bf Note}\ \ 
This is not a mere revision of v1 \cite{Akama},
	where the concept of information condensate was introduced and investigated,
	but it is not a subject of the present paper. 
Here, we advocate a universal information-accumulation mechanism in nature
	due to competition/selection for compositeness, 
	which, we claim, implies a physical law complementary to the entropy-increase law. 
It is an extension of Darwin's natural selection in its literal meaning,
	i.e.\ selection itself without self-replication and heredity, 
	which are strategies selected and accumulated as information later.
The selection is extended to include physical and inner selectors,   	
	and its object is extended from living beings to general composites. 
Its theoretical aim is extended from the origin of species to that of information accumulation in general.
We take it as a physical law because it is fundamental and universal in nature.

It is logically established on the basis of Bateson's fundamental definition of information, 
	``a difference which makes a difference", 
	which is widely accepted today.
We call the made ``difference" its {effect}.
Then, {\it selection for an effect makes information for the effect}.
All the composites formed with information, compete for existence, 
	and are sifted by various selectors. 
Consequently, information is accumulated successively together with specific surviving composites.
To confirm universality of the mechanism, 
	we showed that it works well even in simplest composites such as an atom and a molecule.
Information is indeed a fundamental concept in nature as space-time, matter and energy are!

It would be interesting to investigate competition -- information structures of
	(i)~physical and chemical composites, 
	(ii)~prebiotic and biochemical composites, 
	(iii)~parts, individuals, and groups of living organisms, 
	(iv)~self-organizing systems,
	(v)~geological and meteorological objects, 
	(vi)~psychological entities (cognitions, emotions, etc.), 
	(vii)~social entities, 
	(viii)~cultural entities (technologies, sciences, languages, etc.), 
	(ix)~celestial entities, 
	(x)~artificial entities inclusive of interplays with human, 
	(xi)~hard and soft entities in information sciences, 
	and (xii)~all other composites.
They are all in competition borderlessly, develop strategies for duration, and accumulate information. 
It will provide universal foundations of chemical, biological, cultural, linguistic and cosmological evolutions. 
It would be interesting to study the evolution of strategies for duration and of information itself.


\begin{thebibliography}{99}
\bibitem {literal}
This means selection itself without the implicit premises of
	self-replication, heredity etc., which are strategies 
	selected and accumulated as information later \cite{Darwin}.
\bibitem{intention}
We require no ``intention" here, but as a result they endure 	
	and the information enabling them endures.  
This reasoning is just the same as that in the evolution theory \cite{Darwin}.
The intention etc.\ are based on other information later accumulated to reinforce the functions.
\end{thebibliography}
\end{document}